\documentclass{llncs}
\usepackage{graphics}
\usepackage{graphicx}
\usepackage{amsmath}
\usepackage{amssymb}
\usepackage{graphics}
\usepackage{graphicx}
\usepackage{amstext}
\usepackage{bbold}

\newcommand\rinf{{\underline{\rho}}}
\newcommand\rsup{{\overline{\rho}}}
\newcommand\oinf{{\underline{\omega}}}
\newcommand\osup{{\overline{\omega}}}

\newcommand\ind[1]{\mathbb{1}_{#1}}

\begin{document}

\mainmatter  

\title{Sender and Receiver Energy Consumption in a Sensor Network }

\author{ J.M. Fourneau, F. Quessette}

\institute{
DAVID, Univ. Paris-Saclay, UVSQ, Versailles France\\
\email{jean-michel.fourneau@uvsq.fr} \\
\email{Franck.Quessette@uvsq.fr}
}
\maketitle

\begin{abstract}
We consider a new type of model with energy packets and data packets where the transmission
of a data packet requires energy on both the sender and the receiver nodes. 
Energy packets is a discrete number of Joules representing the quantum of energy needed to send and receive 
a data packet. Both types of packets are stored
in queues. The energy packet queue models a battery. 
Without energy on the sender, the 
emission is delayed until energy is available.  When a sent packet arrives on a receiver which does not have 
enough energy, it is lost. This mechanism implies a new 
complex synchronization between several queues. Despite this complexity, 
we prove that under some classical assumptions  
the steady-state distribution of the Markov chain has a product form solution. We state sufficient conditions 
for ergodicity and we also prove the convergence and the correctness of a numerical algorithm 
to compute the values of the flows. 
\end{abstract}

\section{Introduction} 

Energy Packet Networks (EPNs) were recently introduced  by Gelenbe and his colleagues 
(see for instance \cite{GeMa15,GeCe16,FMBa16,DoFo19} and references therein). 
They can model the flow of intermittent sources of energy
like batteries and solar or wind based generators and study their interactions with 
IT devices consuming energy like sensors, cpu, storage systems and networking elements. 
The key property of this interaction is that one must use energy 
to process the data. Without energy the jobs are delayed or the packets are lost.
The key idea of EPNs is to represent energy with discrete units called Energy  Packets (EPs). 
Since the EPs are produced by an intermittent source of energy (typically solar panels), 
the flow of EPs is associated with some random processes. The principe of the model is the following: 
EPs are consumed  by some
devices after some random duration to perform requested  works or 
can also be stored in a battery from which they can also leak after a random delay: an EP in these models was 
the exact amount of energy needed to process a job. Some queues, denoted as EP queues will store the EPs while 
Data queues are used to store the Data packets. Here we assume that 
there is a one to one matting between the EP queues and the DP queues.     
Data packets (DP) are used  to represent the data to process. 
The interaction between DP and EP is a simultaneous departure of one EP and one DP, a feature which was non conventional in queueing theory. In most papers on Energy Packet networks
(see for instance \cite{Gele12,GeZh19}), 
the main result is a sufficient condition on flow equation to obtain a product-form invariant distribution
(a notable exception being \cite{AbGe16}, where the authors model such a system with diffusion
processes).  Assuming ergodicity 
and the existence of the flow equation solution inside the stability domain, the authors obtain that the steady-state distribution 
has this multiplicative closed form. However both the ergodicity and the existence of the solution are rarely considered.  
The existence of the solution was proved under very strict sufficient conditions: the hyper-stability 
condition  \cite{GeSc92} or the DAG topology assumption \cite{DoFo19}. 

Here we model data transfer in a wireless IOT. Thus we need energy to send the packet and receive it, a
problem which was not studied so far as it implies the synchronization of $4$ queues: 
more precisely two Energy queues and two Data queues. Despite this complex 
synchronization, we can prove the the steady-state distribution has a product form distribution if the flow equation has a 
solution which satisfies the stability constraints. We first prove, as in most of the models studied 
so far,  that the existence of a solution of the flow equations which satisfies  summability constraints implies 
the existence of a product form invariant distribution.  
Here, both ergodicity and invariant distribution solution are proved under the assumption of the existence of a 
solution to the flow equation inside the summability domain.  Thus we have to prove sufficient condition 
for such a solution to exist.
The flow equations are fixed point equations and the natural tool to prove the existence of a solution
is Brouwer's fixed point theorem. 
 A technical difficulty appears immediately: the stability condition is an open set
while Brouwer's theorem require a compact set. To overcome this theoretical 
difficulty,  we propose an algorithmic approach which computes upper and lower bounds 
for the solution. Sufficient conditions for the  convergence will then prove the existence of a solution. 

Usually the flow equations arising for G-networks and EPN are a simple iteration until a fixed point solution is 
proved. This approach is in general not proved and may lead to some problem, being not numerically stable. 
Two notable exceptions are   \cite{Four91} where a sandwich algorithm was proposed to find the solution of the
flow equation 
for a network with positive and negative customers. In  \cite{FoQu06}, we present a different approach where to 
avoid numerical problems, the topology of the network (and the flow equations) are transformed by an elimination
process.  The solution that we proposed here is a generalization of   \cite{Four91}.

The following of the paper is as follows. In Section 2, we present the model and the multiplicative solution for the 
invariant distribution and relate it to the solution of a system of non linear flow equations.
In Section 3, we provide some sufficient conditions for ergodicity while in Section 4
we present and prove an algorithm to compute the solution of the flow equation. Due to the non linearity 
of the flow equations in these networks, no proved algorithms were known for a
general topology \cite{DoFo19}. The application of the model to get the loss rates
for both EP and DP and the average sojourn time for collect trees are presented in Section 2 
and some numerical examples in Section 5. 

\section{The model}

We consider a network of $n$ DP-EP cells. Each cell $i$ has a DP-queue to store the Data Packets and an EP-queue 
(a battery) to store the Energy Packets (respectively denotes as DP and EP in the following). Basically the processing 
of a DP requires an EP. Depending of the type of processing it may lead to a delay or to a loss as detailed now.
\begin{itemize} 
\item The DP moves from cell to cell but it requires energy on both the sender and the receiver. More precisely, at cell $i$, 
if there is some EP available, a DP is sent, if there is one available, to cell $j$ with probability $P_{ij}$ and rate  $\mu_i$. 
If DP queue at cell $i$ is empty then the EP is lost. 
At receiver cell $j$, the DP sent by cell $i$ must use one EP from cell $j$ to be received. If such an EP cell is not available
the DP cell is lost (an emitted packet is lost if the receiver is not able to hear it). Otherwise, the data packet enters the DP-queue at cell $j$. 

\item Note that this energy consumption while the Data queue is empty is consistent with the observations on processor
energy consumption (see \cite{MFC21} and references therein). 
\item The EP external arrivals follow  independent Poisson processes with rate $\alpha_i$ at cell $i$. 
\item The DP external arrivals at cell $i$ follow two independent Poisson processes: with rate 
$\lambda_i$ it consumes one energy packet to enter the cell while with rate 
$\psi_i$ it does not use energy.
\item There is some leakage of EP with rate $\beta_i$.
\item Both $EP$ and $DP$ queues have an infinite capacity. 
\item We assume that $P_{ii} = 0$. Indeed, $P_{ii} >0$ implies that 
we consume energy and stay in the same cell and this is not relevant for our model.   
We also assume that the graph of matrix $P$ is connected. It is not needed that it is strongly connected. 
\item $d_i$ is the DP routing probability from cell $i$ to leave the network. This
departure needs energy (1 EP).  
Of course we have: 
\item Finally, there is also a possibility to leave  the network without consuming energy with rate $\phi_i$. 
\end{itemize} 

We consider that the arrivals and the departures may require energy or not to add more flexibility to the model. 

\begin{figure} [hbtp]
\centerline{
\includegraphics[width=8cm]{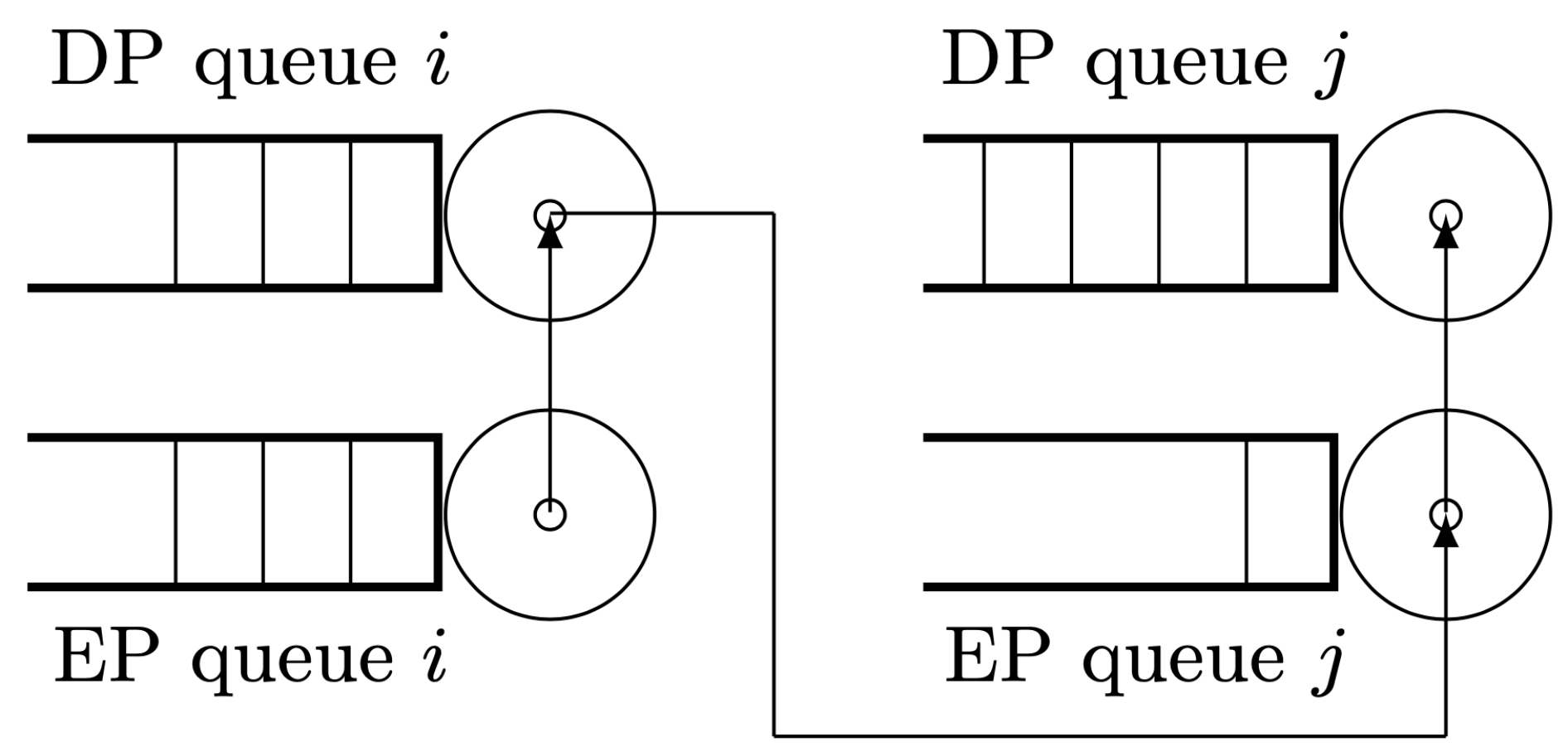}
}
\caption{A successful DP movement consuming one EP to depart cell $i$ and one EP to arrive at cell $j$.} 
\end{figure} 

\subsection{Invariant Distribution} 

Let us first define some particular nodes the network. 

\begin{definition}
A cell is a source if $\lambda_i + \psi_i >0$. Therefore it can receive DP. Similarly a cell is a sink if $\phi_i >0$ or $d_i>0$. 
Data Packets leave the network through a sink cell or being lost during a data transmission. 
\end{definition} 

\begin{definition} [Open network] 
A network of DP and EP cells is open if 
\begin{itemize}
\item For all EP-queue  $i$, $\alpha_i>0$ (i.e. cell $i$ receives energy). 
\item For all EP-queue $i$, we have $\beta_i + \lambda_i + \mu_i >0$ (i.e. the energy at cell $i$ is used). 
\item For all DP-queue $i$, there exists a source cell  $j$ and a directed path from cell $j$ to cell $i$. 
\item  For all DP-queue $i$, there exists a sink cell  $j$ and a directed path from cell $i$ to cell $j$. 
\end{itemize}
\end{definition} 

Let us denote by $x_i$ the number of data packets in DP-queue at cell $i$ and by 
$ y_i$ the number of energy packets in EP-queue at cell $i$. Let 
$\vec{X} = (x_1,x_2,...x_n)$ and $\vec{Y} = (y_1,y_2,...y_n)$ be the number of DP (resp. EP) at each cell. 
Under the assumptions on the network, $(\vec{X},\vec{Y})_t $  is a Markov chain. We now present its invariant distribution.

\begin{theorem}  \label{multiplicative} 
Assume that the flow equations B1,B2 have a solution:
\begin{align*}
\forall i,~~~~ \rho_i = \frac{\psi_i+ \omega_i\lambda_i + \sum_j\rho_j\omega_j\omega_i\mu_jP_{ji}}{\phi_i+\omega_i\mu_i}&&(B1)\\
\forall i,~~~~ \omega_i = \frac{\alpha_i}{\lambda_i + \mu_i + \beta_i + \sum_j \rho_j\omega_j\mu_jP_{ji}}&&(B2)
\end{align*}
then the following distribution $\pi $ is an invariant distribution for the Markov chain. 
\begin{equation} \label{solution}
\pi(\vec{X}, \vec{Y}) = C \times \prod_i \rho_i^{x_i}\omega_i^{y_i}
\end{equation} 
\end{theorem} 

The proof is based on the analysis of the global balance equations. 
 Let $e_i = $ be a  size $n$ null vector except from the $i$th component which is 1.
In the following, all the sums are from 1 to $n$: $\sum_i \lambda_i$ must be read as $\sum_{i=1}^n \lambda_i$. 
We just present here the global balance equation for an arbitrary state $(\vec{X},\vec{Y})$. 
The proof is in appendix A for the sake of readability. 
\begin{equation} \label{gbe} 
\begin{array} {llll} 
\pi(\vec{X},\vec{Y}) & \left[ \sum_i\lambda_i \ind{y_i>0} + \sum_i \mu_i\ind{y_i>0} + \sum_i \alpha_i 
+\sum_i \beta_i \ind{y_i>0} +  \right.   & \left. \sum_i\psi_i + \sum_i\phi_i\ind{x_i>0} \right]  \\
&= \sum_i \pi(\vec{X},\vec{Y}-e_i)\alpha_i\ind{y_i>0}      & \text{Fresh EP arrival}&\\
&+ \sum_i \pi(\vec{X},\vec{Y}+e_i)\beta_i                  & \text{EP Leakage}&\\
&+ \sum_i \pi(\vec{X}-e_i, \vec{Y}+e_i)\lambda_i\ind{x_i>0} & \text{Fresh DP arrival with EP consumption}&\\
&+ \sum_i \pi(\vec{X}+e_i, \vec{Y}+e_i)\mu_i d_i            & \text{DP departure with EP Consumption}&\\
&+ \sum_i \sum_j \pi(\vec{X}+e_i-e_j,\vec{Y}+e_i+e_j)\mu_iP_{ij}\ind{x_j>0} & \text{DP successful movement }&\\
&+ \sum_i \sum_j \pi(\vec{X}+e_i,\vec{Y}+e_i)\mu_iP_{ij}\ind{y_j=0} & \text{Move failed, lack of EP at $j$}&\\
&+ \sum_i \pi(\vec{X},\vec{Y}+e_i)\mu_i\ind{x_i=0} & \text{Mode  failed, lack of DP at  $i$}& \\
&+ \sum_i \pi(\vec{X}-e_i,\vec{Y})\psi_i\ind{x_i>0} & \text{Fresh DP arrival without EP consumption}&\\
&+ \sum_i \pi(\vec{X}+e_i,\vec{Y})\phi_i & \text{DP departure without EP Consumption}& \\
\end{array} 
\end{equation} 

\subsection{Some properties and performance indices} 

It is worthy to remark that the marginal distributions of occupation for EP queues and DP queues
have geometric distributions respectively with rates $\omega_i$ and $\rho_i$. Therefore many formulas
proved for M/M/1 queues are still valid in our model. For instance the average queues size for DP queue 
$i$ is $\rho_i/(1-\rho_i)$  while the average delay is obtained through Little's formula. Similarly the average size of the 
battery is $\omega_i/(1-\omega_i)$ and the probability that battery $i$ is empty is $(1-\omega_i)$. 

A feature of this model is the losses of packet during a transmission when the receiver nodes
do not have energy packets available. The loss rates of DP at queue $i$ is the sum of these rates 
for fresh packets coming from the outside and DP packets migrating between cells:
\[
\lambda_i (1-\omega_i) + (1-\omega_i) \sum_j \rho_j \mu_j \omega_j . 
\]
It is also possible to obtain a relation between DP queue $i$ and EP queue $i$ in isolation as 
established in the following property 
to show the link between energy and performance. 
\begin{property} 
The load at DP queue $i$ is decreasing with the average occupancy of the associated battery. 
\end{property} 
Proof: We consider equation $B2$ and multiply both sides by the denominator. After moving a term from the l.h.s. to the 
r.h.s., we get:
\[
\sum_j \rho_j\omega_j\mu_jP_{ji}  = \alpha_i - \omega_i (\lambda_i + \mu_i + \beta_i) ,
\]
that we substitute in Equation B1. After cancellation of  $\omega_i \lambda_i$, we obtain:
\begin{equation} \label{linkrhoomega}
\rho_i = \frac{\psi_i+ \alpha_i -   \omega_i ( \mu_i + \beta_i) }{\phi_i+\omega_i\mu_i}. 
\end{equation}
Clearly, $\rho_i$ is decreasing with $ \omega_i $. Thus it is also decreasing with the average battery occupancy. 
It is also worthy to remark that when $\omega_i =1$ for all $i$, the network is equivalent to a Jackson network. 
Thus, the model is useful to study how the system performances decrease due to the lack of energy. Furthermore 
the following condition is a necessary condition for stability of the EP queues  for all $i$:
$\alpha_i \leq  \beta_i + \mu_i + \lambda_i 
+ \sum_j \mu_j P_{j,i}$.

\section{Ergodicity} 

Note that the following conditions are only sufficient. One may for instance 
derive ergodic Markov chains associated with networks,
where all the DP are always lost due to the lack of energy in the receiving cells. Clearly, these models are not relevant.  

\begin{lemma} \label{irreductible} 
The Markov chain associated with an open network is irreducible. 
\end{lemma} 
Proof: We prove that for any state $(\vec{x},\vec{y})$ there exists a sequence of transitions with positive probability
from state $(\vec{x},\vec{y})$ to $(\vec{0},\vec{0})$ and another sequence from 
$(\vec{0},\vec{0})$ to $(\vec{x},\vec{y})$. 
Thus the proof is in two parts. 
\begin{itemize}
\item Sequence from $(\vec{0},\vec{0})$ to $(\vec{x},\vec{y})$: in a first subsequence, for all cells, 
we increase the number of $EP$
from $0$ to $y'_i \geq y_i$ such that $y'_i-y_i$ is the energy needed to make the DP enters the network and moves to cell $i$. 
In a second subsequence of events, we let the DP packets enter the sources and move to cell $i$ to reach a number of 
DP equal to $x_i$. The first subsequence of events has a positive probability because $\alpha_i >0$ for all $i$ while the 
second subsequence also has a positive probability as the network is open (assumption 3 for the directed path 
between a source and any
cell $i$). 
\item Sequence from $(\vec{x},\vec{y})$ to $(\vec{0},\vec{0})$: as we  do not assume that $\beta_i >0$, we have to take 
care of how the EP can leave the network. Remember that we have to find a directed path of events which leave 
to a system without energy. Intuitively, in some cases when we have too much EP in the system,  
we must first add some DP  to let them
consume the energy. 

Formally,  we first associate to each cell $i$ a path $path_{j,i}$ from a source $j$ to $i$ ($i$ excluded)  
and a path $path_{i,k}$ 
from $i$ ($i$ excluded) to a sink $k$.  And we compute the number of EP at each cell (say $l(i)$)
needed to empty the whole network
of DP queues using paths $path_{i,k}$ for all $i$. Clearly, we have three cases: 
\begin{itemize}
\item $ y_i = l(i) $. We have nothing to do in that case. 
\item $ y_i > l(i) $. We consider the following subsequence: we provokes the arrivals of $2$ EP at any cell on the path 
$path_{j,i}$ followed by $path_{i,k}$ except EP queue at cell $i$. The source cell at the beginning of  $path_{j,i}$ may only receive 
$1$ EP if the arrival of fresh DP is possible with $\lambda_{j} >0$. Similarly, the sink cell at the end of  $path_{i,k}$ 
may only receive 
$1$ EP if the departure of one DP is possible without energy consumption because $\phi_{k} >0$. Now we provoke the arrival 
of one DP in the source of $path_{j,i} $ (i.e. at cell $j$), 
its transition among $path_{j,i}$, DP queue at cell $i$ and $path_{i,k}$ and finally its departure
from the sink of $path_{i,k}$ (i.e. cell $k$). 
At the end of this subsequence, the number of DP is the same as before in the
whole network, the number of EP is the same as before in the
whole network except in EP queue $i$ where it is decreased by $2$. Indeed a DP need two EP: one at the arrival
and one at the departure to cross a cell. Repeating this sequence leads to the first or the third case depending on the 
parity of $ l(i) - y(i) $. 

\item  $ y_i < l(i) $. We provoke $l(i) - y_i $ arrivals of EP. This subsequence of events has a positive probability as
$\alpha_i >0$. At the completion of this sequence, we have $y(i) = l(i)$ as in the first case. 
\end{itemize} 
 Thus after this subsequences, we have $ y_i = l(i) $ for all cell $i$. We just have to provoke the movements of the 
 DP among paths $path_{i,k2}$. By construction, at the end of this sequence, the network will be empty and all the transitions 
 in the sequence have a positive probability. 
\end{itemize}

\begin{theorem}
Consider an open network of cells, if the solution of the flow equation exists and is such that $\rho_i < 1$ 
and $\omega_i <1$ for all cell $i$, then the Markov chain is ergodic and the steady-state distribution has a multiplicative
form given in Equation \ref{solution} . 
\end{theorem} 
Proof: Theorem \ref{multiplicative} establishes that the distribution 
is an invariant distribution. The conditions on $\rho_i $ and $\omega_i $
proves that this distribution is summable and Lemma \ref{irreductible} establishes that under the open assumption the chain is
irreducible. Applying the results on ergodicity in Bremaud \cite{Brem99}, chapter 8, 
we have that the Markov chain is ergodic and that the invariant
distribution  is the steady-state distribution.

\section{Numerical algorithm}

First remark that if the topology of the directed graph associated with $P$ is Acyclic (i.e. a Directed Acyclic Graph or DAG) 
we can use the topological order associated with the DAG to solve the flow equation in that order as follows:
\begin{enumerate} 
\item change the labels of  the DP queues and the EP queues to use the topological order. 
After this relabeling, we have that $P_{ij} = 0$ if $j<i$. 
\item Solve the equation for $i=1$: 
\begin{align*}
\omega_1 = \frac{\alpha_1}{\lambda_1 + \mu_1 + \beta_1} \\
 \rho_1 = \frac{\psi_1+ \omega_1\lambda_1 }{\phi_1+\omega_1\mu_1} 
\end{align*}
\item iterates for the next values of $i$, knowing that the terms in the summations are computed in the previous iterations
due to the topological order.
\begin{align*}
\rho_i = \frac{\psi_i+ \omega_i\lambda_i + \sum_{j<i} \rho_j\omega_j\omega_i\mu_jP_{ji}}{\phi_i+\omega_i\mu_i}\\
\omega_i = \frac{\alpha_i}{\lambda_i + \mu_i + \beta_i + \sum_{j<i}  \rho_j\omega_j\mu_jP_{ji}}
\end{align*}
\end{enumerate} 
This algorithm does not require any iterations for the solution of the flow equations. 
When the graph contains directed cycles, we must derive another algorithm. 
We proceed in two steps: we first slightly change the system to study a related system where the unknowns evolve in a
compact set. After proving some sufficient conditions of convergence for this new system, we establish some relations to the 
solution of the initial system and the solution of the modified system. More precisely, if the solution of the modified system
is on the boundary of the compact set, then no solution of the initial system exists, while if the solution in in the interior, it
is also a solution of system B1 and B2. 
 
 Remember that we want to find a numerical solution for the system of flow equation $B1$, $B2$. 
 Note that we are dealing with numerical algorithms with floating numbers implemented on a computer.
Thus for any floating number $x$ and $y$, $x=y$ must be interpreted as $|x-y|<\epsilon$ with $\epsilon$ 
a suitable value depending of the floating number representation. $x<y$ and $x\leq y$ are defined in a consistent manner.

We slightly change the system of equations, we present an iterative algorithm et we prove that under some numerical
conditions, this algorithm converges to a solution (see Lemma \ref{sufficient})
which is related to the solution of initial system  (see Lemma \ref{same2}). The proofs are omitted for the sake of conciseness. 
The new system is 
\begin{align*} 
\forall i,~~~~ \rho_i = min(1,\frac{\psi_i+ \omega_i\lambda_i + \sum_j\rho_j\omega_j\omega_i\mu_jP_{ji}}
{\phi_i +\omega_i\mu_i})&&(C1)\\
\forall i,~~~~ \omega_i = min(1, \frac{\alpha_i}{\lambda_i + \mu_i + \beta_i + \sum_j \rho_j\omega_j\mu_jP_{ji}})&&(C2)
\end{align*}

\begin{property}
If we assume that $\phi_i >0$ for all $i$, the system is now continuous on $[0,1]^{2N}$ which is a compact subset of 
$R^{2N}$. Therefore one can use Brouwer's theorem to prove that the new system has a fixed point solution. However
this is only a proof of existence and several solutions may exist on the boundary. The uniqueness of the solution in 
$)0,1(^{2N}$ comes from the uniqueness of the solution for an ergodic chain.  
\end{property} 

The solution for  system $C1$, $C2$ is obtained by iteration. At each step we compute an upper and a lower bound of
all quantities. Let us denote by:
\begin{itemize}
\item $\forall i$, $\rinf_i^{(n)}$, (resp. $\rsup_i^{(n)}$) is the lower (resp. upper) value of $\rho_i$ after iteration $n$. 
\item $\forall i$, $\oinf_i^{(n)}$, (resp. $\osup_i^{(n)}$) is the lower (resp. upper) value of $\omega_i$ after iteration $n$. 
\end{itemize}
For all $i$, we use the following initialization: for all $i$, $\rinf_i^{(0)} = \oinf_i^{(0)} = 0$ 
and $\rsup_i^{(0)} = \osup_i^{(0)} = 1$. 
And we clearly have:  $\rinf_i^{(0)} \leq \rsup_i^{(0)}$ and $\oinf_i^{(0)} \leq \osup_i^{(0)}$. 
Then we iterate as follows to obtain the new versions of $\rinf_i^{(n+1)}$, $\rsup_i^{(n+1)}$, $\oinf_i^{(n+1)}$, $\osup_i^{(n+1)}$:
\begin{align*}
\forall i,~~~~ \rinf_i^{(n+1)} &= \min(1,\frac{\psi_i+ \oinf_i^{(n)}\lambda_i + \sum_j\rinf_j^{(n)}\oinf_j^{(n)}\oinf_i^{(n)}\mu_jP_{ji}}{\phi_i+\osup_i^{(n)}\mu_i})\\
\forall i,~~~~ \rsup_i^{(n+1)} &= \min\left(1,\frac{\psi_i+ \osup_i^{(n)}\lambda_i + \sum_j\rsup_j^{(n)}\osup_j^{(n)}\osup_i^{(n)}\mu_jP_{ji}}{\phi_i+\oinf_i^{(n)}\mu_i}\right)\\
\forall i,~~~~ \oinf_i^{(n+1)} &= \min(1,\frac{\alpha_i}{\lambda_i + \mu_i + \beta_i + \sum_j \rsup_j^{(n)}\osup_j^{(n)}\mu_jP_{ji}})\\
\forall i,~~~~ \osup_i^{(n+1)} &= \min\left(1,\frac{\alpha_i}{\lambda_i + \mu_i + \beta_i + \sum_j \rinf_j^{(n)}\oinf_j^{(n)}\mu_jP_{ji}}\right)\\
\end{align*}
When $\phi_i=0$ for some $i$, to avoid a problem with the definition of $\rsup_i^{(1)}$, we set $\rsup_i^{(1)}=1$.
Note that we do not have a problem to define $\rsup_i^{(2)}$ as $\oinf_i^{(1)}>0$ and the denominator is positive. 
The key ideas of the proof are the following (the properties will be proved in the lemmas): 
\begin{itemize} 
\item The sequence $\rinf_i^{(n)} $ is non decreasing  and the sequence   $\rsup_i^{(n)} $ is non increasing. And:
\[
\forall i,~~~0 \leq \rinf_i^{(n)} \leq  \rinf_i^{(n+1)} <  \rsup_i^{(n+1)} \leq \rsup_i^{(n)} \leq 1 ~~~~ (I1)\\
\]
\item Similarly, 
\[
\forall i,~~~0 \leq \oinf_i^{(n)} \leq  \oinf_i^{(n+1)} <  \osup_i^{(n+1)} \leq \osup_i^{(n)} \leq 1 ~~~ (I2)\\
\]
\item And finally, with an $\epsilon < 1$ which will be given later
\[
\max_{i}~\left(\rsup_i^{(n+1)} - \rinf_i^{(n+1)}, \osup_i^{(n+1)} - \oinf_i^{(n+1)} \right)  \leq \epsilon ~
\max_{i}~\left(\rsup_i^{(n)} - \rinf_i^{(n)}, \osup_i^{(n)} - \oinf_i^{(n)} \right) ~~~ (I3).\\
\]
\end{itemize} 

Let us now proceed with the proof of convergence for the algorithm. 
\begin{lemma} \label{monotone} 
The sequences $\rinf_i^{(n)} $ and $\oinf_i^{(n)}$ are  non decreasing  while the sequences  $\rsup_i^{(n)} $ and  
$\osup_i^{(n)} $ are  non increasing. Furthermore we have $\rinf_i^{(n)} \leq \rsup_i^{(n)}$ and $\oinf_i^{(n)} \leq \osup_i^{(n)}$. 
\end{lemma} 
Proof: by induction on $n$. It is  clear that as all the quantities are non negative, we have 
$\rinf_i^{(1)}  \geq 0 $ and 
$\rsup_i^{(1)}  \leq 1 $. Thus due to the initializations proposed we have: 
$\rinf_i^{(1)}  \geq \rinf_i^{(0)}  $ and $\rsup_i^{(1)}  \leq \rsup_i^{(0)}  $ and 
$\rinf_i^{(0)} \leq \rsup_i^{(0)}$. 
Therefore the property is proved for $n=0$. 
The proof is similar for $\osup_i^{(1)}$ and   $\oinf_i^{(1)}$ and it is omitted. Now assuming for an arbitrary $n$ and for all $i$,
we have: 
\[
\rinf_i^{(n)} \leq \rsup_i^{(n)},~ \rm{and} ~\oinf_i^{(n)} \leq \osup_i^{(n)}. 
\]
Thus, when $n>0$, we have $ 0 < \phi_i +  \mu_i  \oinf_i^{(n)} \leq \phi_i + \mu_i \osup_i^{(n)} $, and 
$\psi_i + \oinf_i^{(n)} \lambda_i + \sum_j \rinf_i \oinf_j \oinf_i \mu_j P_{ji}  
\leq \psi_i + \osup_i^{(n)} \lambda_i + \sum_j \rsup_i \osup_j \osup_i \mu_j P_{ji}  $.  
Therefore we get: 
\[
\frac{\psi_i + \oinf_i^{(n)} \lambda_i + \sum_j \rinf_i^{(n)} \oinf_j^{(n)} \oinf_i^{(n)} \mu_j P_{ji}}{\phi_i + \mu_i \osup_i^{(n)} } \leq 
\frac{\psi_i + \osup_i^{(n)} \lambda_i + \sum_j \rsup_i^{(n)} \osup_j^{(n)} \osup_i^{(n)} \mu_j P_{ji}}{\phi_i +  \mu_i  \oinf_i^{(n)}}.  
\] 
Taking the minimum with $1$, on both sides of the relation, we get:
\[
min(1, \frac{\psi_i + \oinf_i^{(n)} \lambda_i + \sum_j \rinf_i^{(n)} \oinf_j^{(n)} \oinf_i^{(n)} \mu_j P_{ji}}{\phi_i + \mu_i \osup_i^{(n)} } ) \leq 
min(1, \frac{\psi_i + \osup_i^{(n)} \lambda_i + \sum_j \rsup_i^{(n)} \osup_j^{(n)} \osup_i^{(n)} \mu_j P_{ji}}{\phi_i +  \mu_i  \oinf_i^{(n)}}).  
\] 
We  recognize easily the next iterates for $\rinf_i^{(n+1)}$ and $\rsup_i^{(n+1)}$,  
and we get the induction for $n+1$: $\rinf_i^{(n+1)} \leq \rsup_i^{(n+1)}$. 
The proof for the induction on $\oinf_i^{(n+1)}$ and $ \osup_i^{(n+1)}$ is similar and it is omitted for the sake of conciseness. 

Note that the bounded convergence theorem proves that all these sequences converge as they are bounded and monotone. 
However this is not sufficient to show that the limits are the same for the lower bounding and upper bounding sequences. 
We will show in the following that such a difficulty may occur.

But  let us turn to a proof of a sufficient  condition for the 
convergence for the iterative algorithm. We first have to introduce some notation
based on the first iterates. 

Clearly, we have $\rinf_i^{(1)} = min(1,\frac{\psi_i}{\phi_i+\mu_i})$, 
$\rsup_i^{(1)} = min(1,\frac{\psi_i+\lambda_i+\sum_j \mu_j P_{j,i}}{\phi_i})$, \\
$\osup_i^{(1)} = min(1,\frac{\alpha_i}{\lambda_i+\mu_i+\beta_i})$,  and  
$\oinf_i^{(1)} = min(1,\frac{\alpha_i}{\lambda_i+\mu_i+\beta_i+\sum_j \mu_j P_{j,i}})$. Then, to simplify the presentation 
of the results, we define 
\[
a_i = \sum_j \mu_j P_{j,i} \oinf_i^{(1)} \rinf_i^{(1)} , ~~ b_i = \mu_i + \lambda_i + \beta_i , ~~ c_{j,i} = \mu_j P_{j,i}. 
\]

\begin{lemma} [Sufficient condition of convergence for $C1$, $C2$] \label{sufficient} 
Assume that for all $i$ we have, for a particular value of $k$, $\rsup_i^{(k)} < 1$ and 
$\osup_i^{(k)} < 1$, then if 
for all $i$, $\frac{2 \alpha_i}{(a_i + b_i) ^2}\sum_j c_{j,i} < 1$ and 
$\frac{\mu_i\left(2\lambda_i + 4 \sum_j c_{ji} \right)}{(\phi_i+ \mu_i \oinf_i^{(1)}) ^2} <1$, 
then we have a geometric convergence for the algorithm. 
\end{lemma} 
Proof: we prove that in condition $I3$, 
$\epsilon = max_i (\frac{2 \alpha_i \sum_j c_{j,i}} {(a_i + b_i) ^2}, 
\frac{\mu_i\left(2\lambda_i + 4 \sum_j c_{ji} \right)}{(\phi_i+ \mu_i \oinf_i^{(1)}) ^2})$. Thus the assumptions  
imply a geometric convergence of the sequences
to the limit. 
See Appendix B for the detailed proof. 
We now turn on the relation between the solutions of system $C1$, $C2$ and the possible solutions 
of system $B1$ and $B2$. 
The first part is rather clear. 
\begin{lemma} \label{same1} If the algorithm converges for system $C1$ and $C2$. Let 
 $(\hat{\rho}_i $, $\hat{\omega}_i)$ be this solution. 
Assume that, for all $i$, $\hat{\rho}_i < 1$ and  $\hat{\omega}_i<1$, then  $(\hat{\rho}_i ,\hat{\omega}_i)$ 
is also a solution for  system $B1$ and $B2$ and it satisfies the stability conditions. 
\end{lemma} 

When the algorithm converges to a point such that $\hat{\rho}_i < 1$ and  $\hat{\omega}_i<1$,
then we have found a solution of our system. Let us now turn to the other cases. We begin with a technical 
result. 
 
\begin{lemma} \label{same2} Assume that the algorithm converges for system $C1$ and $C2$. 
Assume that there exists a positive 
solution ($\hat{\rho}_i $, $\hat{\omega}_i$) to the system $B1$ and $B2$ such that the 
summability constraints hold (i.e. $\forall i$, $\hat{\rho}_i < 1$ and $\hat{\omega}_i < 1$), then the  solution founded 
for system $C1$ and $C2$ is not on the boundary of the domain (i.e. all  the components are strictly smaller than $1$) 
and it is  ($\hat{\rho}_i $, $\hat{\omega}_i$) . 
\end{lemma} 
Proof: By assumption we have for all $i$:
$
\rinf_i ^{(0)} < \hat{\rho}_i < \rsup_i ^{(0)}   ~~~and~\oinf_i ^{(0)} < \hat{\omega}_i < \osup_i ^{(0)}.
$
And by induction: 
$
\rinf_i^{(n)} \leq \hat{\rho}_i \leq \rsup_i ^{(n)}   ~~~and~\oinf_i^{(n)} \leq \hat{\omega}_i \leq  \osup_i ^{(n)}
$
implies that 
$
\rinf_i ^{(n+1)} \leq  \hat{\rho}_i \leq  \rsup_i ^{(n+1)}   ~~~and~\oinf_i ^{(n+1)} \leq  \hat{\omega}_i \leq \osup_i {(n+1)}.
$
Therefore if the limits  given by the algorithm exist, they are equal to $(\hat{\rho}_i, \hat{\omega}_i)$. 

\begin{lemma} 
If the algorithm converges to a solution (say  ($\hat{\rho}_i $, $\hat{\omega}_i$)). Assume the there exists an $i$ such that 
$\hat{\rho}_i =1$ or  $\hat{\omega}_i=1$, then system $B1$ and $B2$ does not have a solution which satisfies
the summability constraints. 
\end{lemma} 
Proof: it is the contrapositive of the previous Lemma. 

But it is not needed to wait until we find a solution for all the cells. 
\begin{lemma} 
As soon as we have found one cell (say $i$)
such that at iteration $n$, we have  $\rinf_i ^{(n)} =  \rsup_i ^{(n) =1}$ or  $\oinf_i ^{(n)} =  \osup_i ^{(n) =1}$
we prove that system $B1$ and $B2$ does not have a solution which satisfies
the summability constraints. 
\end{lemma} 

Finally, we have to consider the cases where 
$\rsup_i^{(n)}$ and then $\osup_i^{(n)}$ are stuck at $1$ and the other sequences do not evolve anymore
(in that cases, we cannot evoke Lemma \ref{sufficient}). We denote this phenomenon as the grey zone. 

\begin{definition} [Grey Zone]
A queue $i$ is said to be in the grey zone at step $n$, if  the algorithm stops at step $n$ and $\rinf_i ^{(n)} < \rsup_i ^{(n)}$ or $\oinf_i ^{(n)} < \osup_i ^{(n)}$.
\end{definition} 

We give in the following section an example of such a network. Furthermore 
if such a queue exists, they are elements of more complex topological structures in the network. 

\begin{lemma}
If the algorithm stops in the grey zone, it exists at least one cycle of queues (according to the routing matrix $P$) that are in the grey zone.
\end{lemma}
Proof: if $\rinf_i ^{(n)} < \rsup_i ^{(n)}$ (or $ \omega$) it means that the values used to compute $\rinf_i ^{(n)}$ and $\rsup_i ^{(n)}$ (or $\omega$) are different
and necessarily come at least from another queue $j$ in the grey zone and a routing $P_{ji} \neq 0$.
In the same way, if it exists $k$ such that $P_{ik} \neq 0$, 
then queue $k$ is in the grey zone. At the end, as the number of queues is finite, there must be a cycle of grey zone queues.

To escape from the grey zone, we suggest a slightly different algorithm based on the previous one.
Assume that the algorithm stopped at step $n$ with queue $i_0$ in the grey zone with 
$\rinf_{i_0}^{(n)} < \rsup_{i_0}^{(n)}$ (or $\omega$ wlog).
\begin{enumerate}
\item Let us note $\forall j$, $\rinf_j ^{(S)} = \rinf_j^{(n)}$, idem for $\rsup_j^{(S)}$, and $\omega$. $S$ stands for START.
\item In the algorithm, change the equation as follows for all  the next $n$: 
\[
\rinf_i^{(n+1)} = \max(\rinf_j^{(S)}, \min(1,\frac{\psi_i+ \oinf_i^{(n)}\lambda_i + \sum_j\rinf_j^{(n)}\oinf_j^{(n)}\oinf_i^{(n)}\mu_jP_{ji}}{\phi_i+\osup_i^{(n)}\mu_i}))
\]
	Similarly add a $\min$ with $\rsup_j^{(S)}$ for the computation of $\rsup_j^{(n+1)}$ and make the same 
	modification for $\omega$.

	Note that the queues with $\rinf_j ^{(S)} = \rsup_j^{(S)}$ (or $\omega$) has already converge and will not change their values in all future iterations.
\item For the queue $i_0$ in the grey zone with $\rinf_{i_0} ^{(n)} < \rsup_{i_0}^{(n)}$,
	set the value of  $\rsup_{i_0}^{(S)}$ not to $\rsup_{i_0}^{(n)}$ as suggested in Step 1 but, at random 
	to a value strictly lower than $\rsup_{i_0}^{(n)}$ 
	and greater than $\rinf_{i_0}^{(n)}$ .
\item Execute the algorithm starting with the $(n+1)$th iteration and the above modifications.
\end{enumerate}

When the algorithm stops again  (say at iteration $m$), one of the three case may arise for queue $i_0$:
\begin{enumerate}
\item $\rinf_{i_0}^{(m)} = \rsup_{i_0}^{(m)} < \rsup_{i_0}^{(S)}$: the queue $i_0$ is no more in the grey zone for $\rho_{i_0}$. 
If we do not find any queue in the grey zone, then the solution is found. 

If one queue is still in the grey zone, restart the previous algorithm with $i_0$ set to this new queue and $n$ set to $m$.

\item $\rinf_{i_0}^{(m)} < \rsup_{i_0}^{(m)} < \rsup_{i_0}^{(S)} $: the queue $i_0$ is still in the grey zone.  Restart 
the algorithm with $n$ set to $m$ and then a new values for the start values, in particular, $\rsup_{i_0}^{(S)}$ chosen at random between $\rinf_{i_0}^{(m)}$ and $\rsup_{i_0}^{(m)}$.
\item $\rsup_{i_0}^{(m)} = \rsup_{i_0}^{(S)}$: change the value of  $\rsup_{i_0}^{(S)}$  at random
for a greater value but still less than $\rsup_{i_0}^{(n)}$ and rerun the algorithm from iteration $(n+1)$.
\end{enumerate}

The algorithm may never stop if each time it reruns, it ends in the third case. In this case, we can not even concluded that a solution exists or not.

\section{Examples} 

\begin{figure}[hbtp]
\centerline{
\includegraphics[scale=0.75]{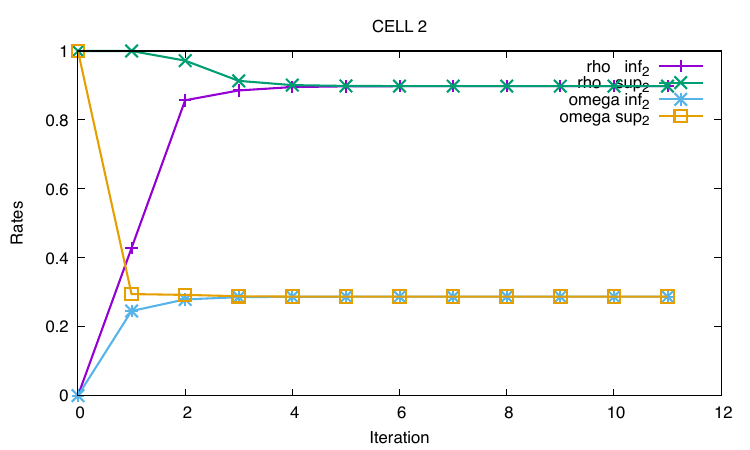} 
}
\caption{Example of convergence for a complete network with 7 cells. We only report the results for Cell $2$.} \label{cell2}
\end{figure}

\begin{figure}[hbtp]
\centerline{
\includegraphics[scale=0.75]{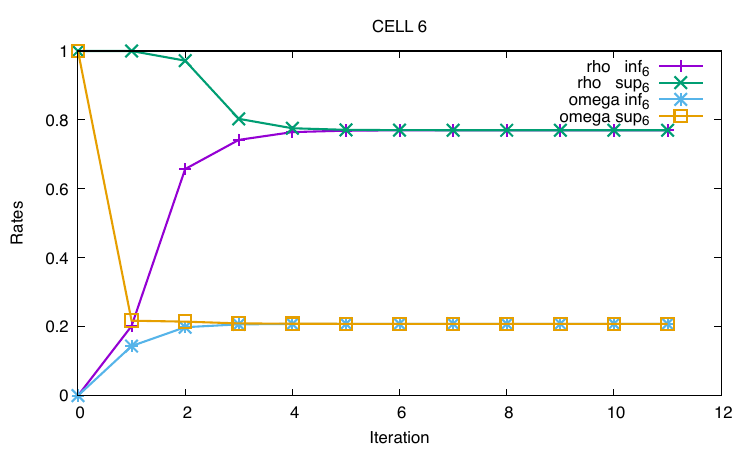} 
}
\caption{Example of convergence for a complete network with 7 cells. Results for Cell $6$.} \label{cell6}
\end{figure}

We present two numerical examples. The first one is a network of $7$ cells with a complete routing matrix. 
We illustrate the convergence process for two cells in Fig. \ref{cell2} and Fig. \ref{cell6}. As the   condition 
of Lemma \ref{sufficient} holds, the convergence is clearly very fast. 

The second example is a small network with a gray zone during the solution procedure. The network consists in two cells
which have the same parameters: $\lambda_i=0$, $\mu_i=3$, $\alpha_i=6.5$, $\beta_i=0.5$, $\phi_i=\psi_i =1$
and $P_{1,2} = P_{2,1} = 1$. After $67$ iterations, the numerical process does not evolve anymore and the final values are: 
$\rinf_i = 0.99999723$, $\rsup_i=1$, $\oinf_i = 0.99999954 $ and $\osup_i =1 $. 

\section{Conclusion} 
In many papers, EPN networks were shown to have a product form steady state distribution if 
a solution for the flow equations exist. Such an assumption is not that simple to prove as the 
equations are neither  linear nor contracting in the general case.  
We hope that this new results will help to develop new algorithms to find these solutions with a more complex
synchronization between queues or stochastic automata network \cite{DaFo09}. These algorithms will be implemented in our 
Markovian Analysis tool XBorne \cite{FAQV16}). 
For instance the load balancing studied in  \cite{BDFo24} models the transfer 
between two DP queueing consuming energy on both the sender and the receiver nodes. We also want to check 
how to generalize the sandwich algorithm in \cite{Four91} to networks of stochastic automata  with Domino
synchronizations  \cite{Four08} which were proved to a multiplicative solution for their steady-state distributions under 
the existence of a fixed point solution for a set of equations on the eigenvectors.

\section*{Appendix A : proof of Theorem. 1} 

We consider the global balance equation (i.e. Eq. \ref{gbe}) and we divide both sides of the equation by 
$\pi(\vec{X},\vec{Y}) $. We assume that the solution is given by the multiplicative form in Eq. \ref{solution}.
After  simplification of the ratio of probabilities, and exchanging indices $i$ and $j$ in the sixth term
of the r.h.s.,  we get: 
\[
\begin{array} {lll} 
\sum_i \lambda_i \ind{y_i>0} + \sum_i \mu_i\ind{y_i>0}  & + \sum_i \alpha_i 
+\sum_i \beta_i \ind{y_i>0} +  \sum_i\psi_i + \sum_i\phi_i\ind{x_i>0} \\
&= \sum_i 1/\omega_i \alpha_i\ind{y_i>0}      \\
&+ \sum_i \omega_i \beta_i                  \\
&+ \sum_i  \omega_i \lambda_i  /\rho_i \ind{x_i>0} \\
&+ \sum_i  \omega_i \rho_i \mu_i d_i            \\
&+ \sum_i \sum_j  \mu_i \omega_i \omega_j \rho_i /\rho_j P_{ij}\ind{x_j>0} \\
&+ \sum_i \sum_j \mu_j \omega_j \rho_j P_{ji}\ind{y_i=0} \\
&+ \sum_i \mu_i \omega_i \ind{x_i=0}  \\
&+ \sum_i \psi_i /\rho_i \ind{x_i>0} \\
&+ \sum_i \rho_i\phi_i  .\\
\end{array} 
\]
We remark that $\ind{y_i=0} = 1 - \ind{y_i>0}$. We substitute into the terms in the r.h.s. and move the negative terms
into the l.h.s:
\[
\begin{array} {lll} 
\sum_i \lambda_i \ind{y_i>0} + \sum_i \mu_i\ind{y_i>0}  & + \sum_i \alpha_i 
+\sum_i \beta_i \ind{y_i>0} +  \sum_i\psi_i \\
&+ \sum_i\phi_i\ind{x_i>0} 
+ \sum_i \sum_j \mu_j \omega_j \rho_j P_{ji}\ind{y_i>0} 
+ \sum_i \mu_i \omega_i \ind{x_i>0}  \\
&= \sum_i 1/\omega_i \alpha_i\ind{y_i>0}      \\
&+ \sum_i \omega_i \beta_i                  \\
&+ \sum_i  \omega_i \lambda_i  /\rho_i \ind{x_i>0} \\
&+ \sum_i  \omega_i \rho_i \mu_i d_i            \\
&+ \sum_i \sum_j  \mu_i \omega_i \omega_j \rho_i /\rho_j P_{ij}\ind{x_j>0} \\
&+ \sum_i \sum_j \mu_j \omega_j \rho_j P_{ji}  \\
&+ \sum_i \mu_i \omega_i   \\
&+ \sum_i \psi_i /\rho_i \ind{x_i>0} \\
&+ \sum_i \rho_i\phi_i . \\
\end{array} 
\]
We exchange indices $i$ and $j$ in the fifth term of the r.h.s. and we can now factorize according to the step functions both
sides of the equation:
\[
\begin{array} {lll} 
\sum_i \ind{y_i>0}  ( \lambda_i +  \mu_i + \beta_i + \sum_j \mu_j \omega_j \rho_j P_{ji})   & + \sum_i (\alpha_i + \psi_i ) 
+ \sum_i (\phi_i + \mu_i \omega_i ) \ind{x_i>0}  \\
&= \sum_i 1/\omega_i \alpha_i\ind{y_i>0}      \\
&+ \sum_i ( \omega_i \beta_i  + \sum_j \mu_j \omega_j \rho_j P_{ji} + \omega_i \rho_i \mu_i d_i + \rho_i\phi_i  + \mu_i \omega_i  )   \\
&+ \sum_i  1/ \rho_i \ind{x_i>0}  (\omega_i \lambda_i  + \psi_i  + \sum_j  \mu_j \omega_j \omega_i \rho_j P_{ji} ). \\
\end{array} 
\]
Due to Relation (B1) the third term of the l.h.s. cancels with the third term of the r.h.s., and due to relation 
(B2), we also simplify the first term of  the l.h.s. with the first term of the r.h.s. to obtain finally: 
\[
 \sum_i \alpha_i + \sum_i \psi_i  =  \sum_i  \omega_i \beta_i  + \sum_i \sum_j \mu_j \omega_j \rho_j P_{ji} + \sum_i \omega_i \rho_i \mu_i d_i + \sum_i \rho_i\phi_i  + \sum_i \mu_i \omega_i ,
 \]
 which can be further simplified taking into account that for all $i$ we have the normalization relation for the routing:
$
 d_i + \sum_j P_{ji} =1$. 
After some algebraic manipulation of indices and using the normalization of the routing probabilities we get:
\begin{equation} \label{flow} 
 \sum_i \alpha_i + \sum_i \psi_i  =  \sum_i  \omega_i \beta_i  + \sum_i \omega_i \rho_i \mu_i + \sum_i \rho_i\phi_i  + \sum_i \mu_i \omega_i .
 \end{equation} 
It remains to prove that this equation is consistent with the flow equations B1 and B2. Consider first Equation B2 on the flow of
EP. Multiply by the denominator and sum up for all $i$:
\[
\sum_i \omega_i \lambda_i + \sum_i \omega_i \mu_i + \sum_i \omega_i \beta_i 
+ \sum_i \sum_j \omega_i \rho_j\omega_j\mu_jP_{ji}  = \sum_i \alpha_i .
\]
Similarly, using the same approach for Equation B1, we get:
\[
\sum \rho_i \phi_i + \sum_i \rho_i \omega_i\mu_i = \sum_i \psi_i + \sum_i \omega_i\lambda_i + \sum_i \sum_j\rho_j\omega_j\omega_i\mu_jP_{ji} .
\]
We add both equations and the terms $\sum_i \omega_i\lambda_i $ and $ \sum_i \sum_j\rho_j\omega_j\omega_i\mu_jP_{ji} $ cancel as they are on both sides of the resulting equation. We finally get:
\[
\sum_i \omega_i \mu_i + \sum_i \omega_i \beta_i  + 
\sum \rho_i \phi_i + \sum_i \rho_i \omega_i\mu_i
 = \sum_i \alpha_i 
+ \sum_i \psi_i  ,
\]
and this is exactly Eq. \ref{flow}. Thus the proof is complete.

\section*{Appendix B : Convergence } 

We will prove that for all $n\geq k$ we have: 
\[
\max_{i}~\left(\rsup_i^{(n+1)} - \rinf_i^{(n+1)}, \osup_i^{(n+1)} - \oinf_i^{(n+1)} \right)  \leq \epsilon ~
\max_{i}~\left(\rsup_i^{(n)} - \rinf_i^{(n)}, \osup_i^{(n)} - \oinf_i^{(n)} \right),\\
\]
with $\epsilon \leq max_i (2\frac{\alpha_i}{b_i^2}\sum_j c_{ji}, 
\frac{\mu_i\left(2\lambda_i + 4 \sum_j c_{ji} \right)}{(\phi_i+ \mu_i \oinf_i^{(1)})^2}  $. 
First remark that, due to Lemma  \ref{monotone} for all $n\geq k$, all the iterates are strictly smaller than $1$. Thus,
\begin{align*}
\osup_i^{(n+1)} - \oinf_i^{(n+1)} && = && \frac{\alpha_i}{b_i + \sum_j c_{ji} \oinf_j^{(n)}\rinf_j^{(n)}} - \frac{\alpha_i}{b_i + \sum_j c_{ji} \osup_j^{(n)}\rsup_j^{(n)}}\\
&& \leq && \frac{\alpha_i\left(\sum_j c_{ji} (\osup_j^{(n)}\rsup_j^{(n)} - \oinf_j^{(n)}\rinf_j^{(n)})\right)}{b_i^2} && \text{using~} b_i+\sum_j c_{ji} ~\rho_j ~ \omega_j \geq b_i\\
&& \leq &&\frac{\alpha_i}{b_i^2}\left[  
\sum_j c_{ji} \left[\rsup_j^{(n)}(\osup_j^{(n)}-\oinf_j^{(n)}) + \oinf_j^{(n)}(\rsup_j^{(n)}-\rinf_j^{(n)})\right]
\right] && \text{adding and subtracting~}\rsup_j^{(n)}\oinf_j^{(n)}\\
&& \leq && \frac{\alpha_i}{b_i^2}\left[ \sum_j c_{ji} (\osup_j^{(n)}-\oinf_j^{(n)}) + \sum_j c_{ji} ~ (\rsup_j^{(n)}-\rinf_j^{(n)})\right] && \text{since~}\rsup_j^{(n)} \text{~and~}\oinf_j^{(n)} \leq 1\\
&& \leq && \left(2\frac{\alpha_i}{b_i^2}\sum_j c_{ji} \right) \times \left( \max_j\left(\osup_j^{(n)}-\oinf_j^{(n)},\rsup_j^{(n)}-\rinf_j^{(n)}\right)\right)&.&\\
\end{align*}

Let us now bound  $\rsup_i^{(n+1)} - \rinf_i^{(n+1)}$:
\begin{align*}
\rsup_i^{(n+1)} - \rinf_i^{(n+1)} && = && \frac{\psi_i + \lambda_i\osup_i^{(n)} + \sum_j c_{ji}  \rsup_j^{(n)}\osup_j^{(n)}\osup_i^{(n)}}{\phi_i + \oinf_i^{(n)}\mu_i}
- \frac{\psi_i + \lambda_i\oinf_i^{(n)} + \sum_j c_{ji} \rinf_j^{(n)}\oinf_j^{(n)}\oinf_i^{(n)}}{\phi_i + \osup_i^{(n)}\mu_i}\\
&& = && \mu_i \frac{ \lambda_i\left( 
				{\osup_i^{(n)}}^2 - {\oinf_i^{(n)}}^2   \right) 
				+ \sum_j c_{ji} \left( \rsup_j^{(n)}\osup_j^{(n)}{\osup_i^{(n)}}^2  
				- \rinf_j^{(n)}\oinf_j^{(n)}{\oinf_i^{(n)}}^2 \right)
				}{(\phi_i+\oinf_i^{(n)} \mu_i)(\phi_i + \osup_i^{(n)} \mu_i)}.\\
\end{align*}
We first provide some simple  bounds for some of the quantities appearing in the equation. 
\begin{itemize} 
\item ${\osup_i^{(n)}}^2 - {\oinf_i^{(n)}}^2 \leq  2 ({\osup_i^{(n)}} - {\oinf_i^{(n)}} )$ , 
\item $\rsup_j^{(n)}\osup_j^{(n)}{\osup_i^{(n)}}^2  - \rinf_j^{(n)}\oinf_j^{(n)}{\oinf_i^{(n)}}^2 \leq 
4  max_m\left(\osup_m^{(n)} -\oinf_m^{(n)},\rsup_m^{(n)} -\rinf_m^{(n)} \right)$, 
\item and finally, $(\phi_i+\oinf_i^{(n)} \mu_i)(\phi_i + \osup_i^{(n)} \mu_i) \geq 
(\phi_i+\oinf_i^{(1)} \mu_i)^2$. 
\end{itemize}

Going back to the proof, we obtain after substitution:
\begin{align*}
\rsup^{(n+1)} - \rinf^{(n+1)} && \leq && 
\frac{\mu_i\left(2\lambda_i + 4 \sum_j c_{ji} \right)}{(\phi_i+ \mu_i \oinf_i^{(1)} )^2} \times \max_j\left(\osup^{(n)}_j-\oinf{(n)}_j,\rsup{(n)}_j-\rinf{(n)}_j\right). 
\end{align*}

\bibliographystyle{unsrt} 
\bibliography{Borne} 

\end{document}